\documentclass[aps, superscriptaddress, twocolumn, notilepage, floatfix, pra, nofootinbib, longbibliography]{revtex4-2}
\usepackage{amsmath,amssymb,amsthm}
\usepackage{enumitem}

\usepackage{xcolor}

\usepackage{lipsum}

\usepackage{braket}
\usepackage{hhline}

\usepackage[pdftex]{graphicx}
\usepackage{graphics}
\usepackage{mathrsfs}
\usepackage[colorlinks, breaklinks, urlcolor={blue}, linkcolor={red}, citecolor={blue}]{hyperref}
\usepackage{array, makecell}
\usepackage{type1cm}
\usepackage{lettrine}
\usepackage[english]{babel}
\usepackage{lmodern}
\usepackage{microtype}
\usepackage{booktabs}
\usepackage[T1]{fontenc}
\usepackage[boxed, vlined]{algorithm2e}
\usepackage{caption}

\usepackage{subfig}
\usepackage{float}

\usepackage{caption}
\newcommand{\op}[2]{|#1\rangle \langle #2|}

\newcommand{\ov}[1]{\overline{ #1}}

\newcommand{\madhav}[1]{{\color{black} {#1}}}

\usepackage{tikz}
\usetikzlibrary{quantikz}

\begin{document}
\title{Compilation of algorithm-specific graph states for quantum circuits}
\begin{abstract}
   We present a quantum circuit compiler that prepares an algorithm-specific  graph  state  from  quantum  circuits described in high level languages, such as Cirq and Q\#.  The computation \madhav{can then be} implemented using a series of non-Pauli measurements on \madhav{this} graph state.  By compiling the graph state directly instead of starting with a standard lattice cluster state and preparing it over the course of the computation, we are able to better understand the resource costs involved and eliminate wasteful Pauli measurements on the actual quantum device. Access to this algorithm-specific graph state also allows for optimisation over locally equivalent graph states to implement the same quantum circuit.  
   \madhav{The compiler presented here finds ready application in measurement based quantum computing, NISQ devices and logical level compilation for fault tolereant implementations.  }
\end{abstract}
\date{\today}
\author{Madhav Krishnan Vijayan}
\affiliation{Centre for Quantum Software and Information, University of Technology Sydney, Sydney, NSW 2007, Australia}
\author{Alexandru Paler}
\affiliation{Aalto University, 02150 Espoo, Finland}
\author{Jason Gavriel}
\affiliation{Centre for Quantum Software and Information, University of Technology Sydney, Sydney, NSW 2007, Australia}
\affiliation{Centre for Quantum Computation and Communication Technology}
\author{Casey R.~Myers}
\affiliation{School of Computing and Information Systems, Faculty of Engineering and Information Technology, The University of Melbourne, Melbourne VIC 3010, Australia}
\affiliation{Silicon Quantum Computing Pty Ltd., Level 2, Newton Building, UNSW Sydney, Kensington, NSW 2052, Australia}
\author{Peter P.~Rohde}
\affiliation{Centre for Quantum Software and Information, University of Technology Sydney, Sydney, NSW 2007, Australia}
\author{Simon J.~Devitt}
\affiliation{Centre for Quantum Software and Information, University of Technology Sydney, Sydney, NSW 2007, Australia}
\maketitle

\section{Introduction}

\madhav{The circuit model is ubiquitous in the field of quantum computing and is used to represent and analyse implementations of various quantum algorithms. However this is not a unique choice and other models also exist for instance that of measurement based quantum computing (MBQC) \cite{raussendorf_one-way_2001, nielsen_cluster-state_2006}.} In MBQC, one starts with a standard square lattice cluster state, \madhav{which is a certain type of quantum graph state,} and performs single qubit measurements to teleport the logical quantum state through the lattice. \madhav{Gates are applied to the state by the choice of measurement basis and corrections due to the probabilistic nature of measurement outcomes are handled by dynamically adapting the basis of future measurements based on past outcomes \cite{raussendorf_measurement-based_2003} as well as classically tracking the Pauli frame \cite{paler_software-based_2014}.} Pauli basis measurements are sufficient to perform Clifford operations, however to perform non-Clifford operations, \madhav{additional rotated basis measurements are required. There have been recent efforts to build concrete methods for algorithm development in MBQC like software frameworks such as \cite{evans_mcbeth_2022, zhang_compilation_2022}}. 

We present a compiler which employs an alternate workflow where instead of starting with the lattice cluster state, we prepare an algorithm-specific graph state. The latter is essentially a local Clifford equivalent graph state of the system after all the Clifford gates in the circuit have been applied. Once the algorithm-specific graph state is prepared, the computation is carried out by performing \madhav{non-Pauli measurements} on the qubits as before. In this manner we classically perform the Clifford part of the computation efficiently \cite{gottesman_heisenberg_1998, aaronson_improved_2004} and leave only the non-Clifford part to the quantum device. \madhav{Specifying both an algorithm-specific graph state and a measurement procedure completely determines the quantum computation leaving us free to discard the circuit model completely and treat the computation as an instruction set to prepare a graph state and perform a sequence of measurements. This is conceptually similar to stabilizer simulators that use graph states as an underlying representation \cite{anders_fast_2006}, however we demonstrate a compilation framework that accommodates arbitrary circuits and not just  Clifford circuits. 

In this framework, questions of optimal implementations can be addressed in terms of graph optimisation in local Clifford orbits (see section  \ref{sec:resource-optimisation}) opening new avenues for physical architecture designs. The algorithm-specific graph state also naturally represents the connectivity requirements for implementing a particular algorithm allowing us to determine the minimum device resources needed. While from the perspective of compilation we require knowledge of both the algorithm-specific graph state as well the measurement procedure to implement the computation, for the purposes of algorithm benchmarking and quantum resource estimation, it sufficient to know the graph state and the number of non-Pauli measurements required. Another application of algorithm-specific graphs is for logical level compilation for  a fault tolerant error correction code. By encoding the computation in a graph state, we dispense with the requirement of performing complex state evolution in an error corrected code and only require the generation of the graph state followed by logical qubit measurements.} 

To perform all the Clifford operations first, we compile a given quantum circuit described in high level languages, such as Cirq and Q\#, in the so called ICM (Initialisation-CNOT-Measurement) form \cite{paler_fault-tolerant_2017, herr_lattice_2017}, the details of which are given in Sec.~\ref{sec:ICM}. The variation of the ICM form we use allows for any circuit to be implemented \madhav{by a series of Clifford gates followed by non-Pauli measurements.} The \madhav{quantum state prepared by the circuit prior to measurement is a stabilizer state and hence can alternatively be generated from an algorithm-specific graph state by using appropriate local Clifford corrections \cite{van_den_nest_graphical_2004}. To determine this aglorithm-specific graph state, we simulate the Clifford gates directly using well known stabilizer simulation techniques \cite{aaronson_improved_2004, gidney_stim_2021} and we employ a graph conversion algorithm which computes a locally equivalent graph state given any stabilizer state (see section \ref{sec:graph_conversion})}. It is worth pointing out here that there exists an interesting parallel to our techniques in the literature of circuit optimisation using the ZX calculus \cite{coeckeInteractingQuantumObservables2011}. The authors of \cite{backensThereBackAgain2021} for instance are able to remove nodes in the ZX diagram which represent Clifford gates by appropriate rewrite rules. Determining whether there is a exact mathematical correspondence between these two techniques is left to future work.

This paper is structured as follows: In section \ref{sec:prelims} we detail some preliminaries about the stabilizer formalism, graph states and their representation as a tableau. Section \ref{sec:mbqc} gives a brief introduction to measurement based quantum computing. In section \ref{sec:graph_conversion} we describe the stabilizer to graph state conversion and present an example. Then in section \ref{sec:ICM} we review the ICM formalism and show how to decompose an arbitrary quantum circuit into a Clifford block initialisation followed by non-Pauli measurements. We explicitly show this workflow for the controlled-$V^{\dagger}$ and the Toffolli gates in section \ref{sec:algo_examples}. In section \ref{sec:resource-optimisation} we discuss the implications of the local equivalence of graph states for resource optimisation \madhav{and in section \ref{sec:compat_arch} we comment on the utility of the compiler for optical quantum computing, NISQ devices and fault tolerant implementations.} A technical overview of our implementation of the compiler is given in section \ref{sec:tech_overview}. Finally, we conclude and discuss future work in section \ref{sec:conclusion}.

\textcolor{black}{Our implementation of the compiler can be found at \url{https://github.com/QSI-BAQS/Jabalizer}. The source code for all the examples used in this paper can be found on the branch ``manuscript''.}

\section{Preliminaries} \label{sec:prelims}

Graph states are at the foundation of our compiler, and are an important subclass of stabilizer states widely used in measurement based quantum computing \cite{raussendorf_one-way_2001}. Stabilizer states and graph states are related concepts and, in the following, we offer a brief review.

\subsection{Stabilizer Formalism}

Stabilizer states which were initially introduced as \madhav{a formal description of} quantum error correction codes \cite{bennett_mixed-state_1996, calderbank_quantum_1997,gottesman_class_1996} have been extensively studied and find wide application \madhav{ \cite{gottesman_introduction_2009, calderbank_quantum_1997, bombin_structure_2014}. } We will now give a brief review of the mathematical formalism of stabilizer states, their representation and connection to graph states. The interested reader is referred to Sec.~10.5.1 of \cite{bib:NielsenChuang00} for a through treatment.

\subsubsection{Stabilizer states}

\madhav{Operations which map the set of stabilizer states to itself constitute the Clifford group of operations} and the well known Gottesman-Knill theorem proves that \madhav{all such operations can} be efficiently simulated on a classical computer \cite{gottesman_heisenberg_1998}. 

The Pauli group $\mathcal{G}_n$ of $n$ qubits is the set of all $n$-fold tensor product combinations of $\{\pm1, \pm i, \} \times \{I, X, Y, Z\} $, where $I$ is the single qubit identity operator and $X$, $Y$, $Z$ are the single qubit Pauli operators, respectively. We say that a commuting subgroup $\mathcal S_n$ of $\mathcal G_n$ stabilises a vector space $\mathcal V$ if for all elements $s \in \mathcal S_n$ and $ v \in \mathcal{V}$, $sv = v $. These stabilizer groups will consist of Pauli group elements with coefficients +1 and -1 only since they must satisfy $s^2 = I^{\otimes n}$. A concise way of representing these subgroups is by specifying a set of generators of the subgroup $S = \{s_1,s_2,\dots,s_k\}$ such that any element of the group can be represented as products of these generators. For a $2^n$ dimensional system, specifying $n$ independent generators uniquely specifies a  stabilizer state. The set of operations that map the set of all stabilizer states to itself is called the Clifford group of operations. If we only use Clifford operations and measurements in our quantum circuit, we can completely describe the evolution of the state vector classically by keeping track of the stabilizer generators of the state and their evolution under Clifford operations \madhav{ \cite{aaronson_improved_2004, anders_fast_2006}}. 

\subsubsection{Evolving stabilizer states under Clifford operations}
State evolution of stabilizer states can be performed in the Heisenberg picture by conjugating the stabilizers with the unitary operations acting upon the state since for the evolution
\begin{align*}
    \ket{\psi^{\prime}} = U \ket{\psi},
\end{align*}
we can equivalently write,
\begin{align*}
    \ket{\psi^{\prime}} = U s_i \ket{\psi} = U s_i U{^\dagger}\ U \ket{\psi} = s^{\prime}_i \ket{\psi^{\prime}},
\end{align*}
where $s^{\prime}_i$ stabilizes the evolved state $\ket{\psi^{\prime}}$. The efficiency of classically simulating stabilizer circuits in this way was improved upon by \cite{aaronson_improved_2004} using the so-called CHP approach which tracks both stabilizers and anti-stabilizers (anti-stabilizers of a state along with the stabilizers span the entire Pauli group $\mathcal G_n$), improving the performance of measurements within this model. We will confine our discussion to using just the stabilizer generators for simplicity since both methods are conceptually equivalent.

\subsection{Tableau representation}
\label{sec:tableau}
Using the relations $X^2 = Z^2 = I$ and $ZX = iY$, any stabilizer $s \in \mathcal S_n$ can be thought of as tensor products of terms $X^xZ^z$  and a phase $(-1)^p$, where $x,z,p \in \{0,1\}$ are binary variables. This allows us to represent the generators $S$ as rows in a $n \times (2n+1) $ binary matrix
\begin{align}
\begin{bmatrix}\label{eq:tableau}
x_{11} & x_{12}  & \dots &x_{1n} &z_{11}  &z_{12}  &\dots  &z_{1n} &p_{1} \\
x_{21} & x_{22}  & \dots &x_{2n} &z_{21}  &z_{22}  &\dots &z_{2n} &p_{2} \\
  &    &  & &\vdots  &   &   &  \\
x_{n1} & x_{n2}  & \dots &x_{nn}  &z_{n1}  &z_{n2}  &\dots &z_{nn} &p_{n}
\end{bmatrix}.
\end{align}
Such a representation is known as a stabilizer tableau. For a stabilizer $s \in S$ we refer to its corresponding row in the tableau, not including the final phase column, as $r(s)$. The tableau representation has the useful property that for two stabilizers $s_i, s_j \in S$, $r(s_i) + r(s_j) = r(s_is_j)$, where the addition is modulo 2. This means that we can replace any row in the tableau with its sum with another row if we correct for the phases, since if $s_i$ and $s_j$ are a valid stabilizer, so too is $s_is_j$. The phase of the product $s_i s_j$ can be efficiently computed by taking the product of the stabilizers term by term and multiplying together all the resulting phases.

\subsection{Graph states}

Given a mathematical graph defined through a vertex set $V$ and an undirected, unweighted edge set $E$, the graph state is prepared by initialising qubits in the $\ket{+}$ state for each vertex and applying a controlled-$Z$ operation between qubits with a shared edge. Equivalently, they are defined through a canonical stabilizer set of the form
\begin{align}\label{eq:graph_state_def}
    s_i = X_i\prod_{j \in n_i} Z_j
\end{align}
for all vertex labels $i$, where $n_i$ is the edge neighbourhood of the $i^\mathrm{th}$ vertex in the graph. For a graph state, with a diagonal $X$-block (i.e. identity matrix), the $Z$-block in the tableau representation shown in Eq.~\eqref{eq:tableau} will correspond to the adjacency matrix of the \madhav{abstract classical graph.} Thus for a graph state the tableau will be of the form $[I_n|A|p\,]$, where $A$ is the graph adjacency matrix and $p$ is a column vector of phases which are by convention all +1.

\section{Measurement based quantum computing}
\label{sec:mbqc}

This section is a brief introduction to measurement based quantum computing. It is not our intention to be comprehensive and the reader is referred to \cite{ raussendorf_measurement-based_2003, raussendorf_quantum_2012, raussendorf_computational_2002} for further details. We will confine our discussion to elements that have a direct relevance to algorithm-specific graph compilation. 

In MBQC the primary computational resource is a cluster state which is a special kind of graph state. The edges in a cluster state are confined to be between nearest neighbour nodes and the nodes themselves are arranged in a regular lattice structure. We will focus on the canonical example of the the 2D square lattice in which MBQC is known to be universal. Note that other lattice structures are possible and 3D lattices in particular allow for fault tolerant topological codes \cite{raussendorf_topological_2007}. Once the cluster state is prepared the entire computation is carried out by a series of single qubit measurement rounds, with the measured observable for each qubit depending on the measurement outcomes from previous rounds. Once all the measurements are done, the result of the computation can be determined by classical post processing. We now look at the details of how a universal set of gates (and thus any quantum circuit) can be implemented using MBQC.

\subsection{Gate implementation}

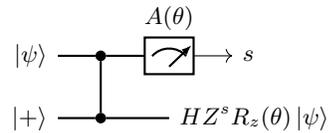
\begin{figure}
\centering
\begin{tikzpicture}
	\node[scale=1]{
		\begin{quantikz}
			\lstick{$\ket{\psi}$}   & \ctrl{1}  &\meter{$A(\theta)$} \arrow[r]  & \rstick{$s$}\\
			\lstick{$\ket{+}$}  & \control{}  &\qw  \rstick{$HZ^sR_z(\theta)\ket{\psi}$} 
		\end{quantikz}
	};
	\end{tikzpicture}
\caption{Gate teleportation circuit: Here $R_z(\theta) = e^{-i\theta Z/ 2 }$ and measuring the observable $A(\theta) = R_z^{\dagger}(\theta) X R_z(\theta)$ on the first qubit has the effect of teleporting the state $R_z(\theta)\ket{\psi}$ to the next site with the addition of a Hadamard gate and Pauli Z correction depending on the measurement outcome.}
\label{fig:gate_teleportation}
\end{figure}

The primary driver of circuit simulation on a cluster state is a teleportation protocol which allows us to perform single qubit rotations. An arbitrary single qubit state can be encoded into a linear cluster using the Controlled-Z operation as shown in figure~\ref{fig:gate_teleportation}.
\noindent Measuring the observable $A(\theta) = R_z^{\dagger}(\theta) X R_z(\theta)$, where $R_z(\theta) =  e^{-i\theta Z/ 2 }$, on the leftmost qubit has the effect of teleporting the state $\ket{\psi}$ to the next site with the addition of a Hadamard gate and Pauli Z correction depending on the measurement outcome. Note that it is often convenient to represent $A(\theta)$ as the z-rotated X observable $\cos\theta X - \sin\theta Y$. Repeating the gate teleportation circuit in figure~\ref{fig:gate_teleportation} twice with measurements angles $\theta_1$ and $\theta_2$ and outcomes $s_1$ and $s_2$, respectively, leads to the output state 
\begin{align}
\label{eq:teleport_output}
    \ket{\psi_{\text{out}}}   &= HZ^{s_2}R_z(\theta_2)HZ^{s_1}R_z(\theta_1)\ket{\psi} \notag \\
                &= X^{s_2}R_x(\theta_2)Z^{s_1}R_z(\theta_1)\ket{\psi} \\
                &= X^{s_2} Z^{s_1} R_x((-1)^{s_1}\theta_2)R_z(\theta_1)\ket{\psi}, \notag
\end{align}
where in the second line we have commuted the Hadamard operator on the left through to meet the one on the right which converts the Z correction and rotation to an X correction and rotation with $R_x(\theta) =  HR_z(\theta)H$. In the last line we commute the Z correction to the left using the relation $e^{i\theta X}Z = Ze^{-i\theta X }$. It is straightforward to see that repeating this procedure we are able to implement a series of alternating $X$ and $Z$ rotations, three of which is sufficient to implement any single qubit unitary gate.

There are two unwanted effects due to the measurement dependent Pauli corrections: 1) there is a Pauli operator of the form $X^{f_x(\vec s)}Z^{ f_z(\vec s) }$, where $f_x(\vec s)$ and $f_z(\vec s)$  are functions of the binary measurement outcome vector $\vec s$ \madhav{(here $\vec s = (s_1, s_2)^T$)}; and 2) The $R_x$ and $R_z$ rotation angles are flipped from $\theta_i$ to $-\theta_i$ if an odd number of Z or X operators are commuted through it, respectively.

We do not need to do any active correction for the first effect since at the end of the circuit we are going to measure in the computational Z basis. The presence of a Z operator has no effect on the Z measurement and the presence of an X operator induces a relabeling between 0 and 1, which can be accounted for in post-processing as long as we keep track of all the measurement outcomes.

\begin{figure}
  \centering
  \subfloat[a][Single qubit rotation]{\includegraphics[width=0.7\linewidth]{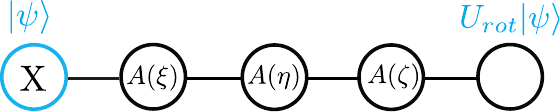} \label{fig:sqr_cluster}} \\
  \subfloat[b][CNOT gate]{\includegraphics[width = 0.75\linewidth]{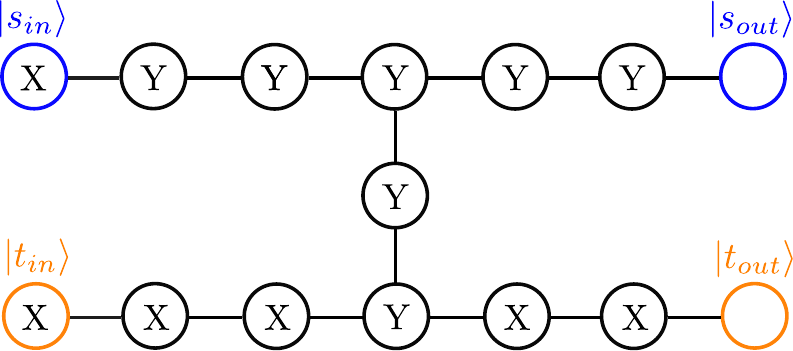} \label{fig:cnot_cluster}}
  \caption{(a) An arbitrary single qubit unitary in the Euler representation $U_{rot} = e^{-i\frac{\zeta}{2 } Z} e^{-i\frac{\eta}{2 } X} e^{-i\frac{\xi}{2 } Z} $ is implemented by the linear cluster shown above by measuring the first qubit in Pauli X basis and the other three qubits in the $A(\theta) = \cos \theta X -  \sin \theta Y$ basis, with $\theta$ being $\xi, \eta$ and $\zeta$, respectively, for the second, third and fourth qubits from the left. Based on previous measurement outcomes, $A(-\theta)$ could me measured instead of $A(\theta)$. (b)  A CNOT gate can be implemented between a source qubit and a target qubit using the 15 qubit cluster state shown and simultaneous Pauli measurements as indicated on each qubit.} \label{fig:gate_cluster.}
\end{figure}

As for the second effect, as seen in equation \eqref{eq:teleport_output}, whether $R_x(\theta)$ or $R_x(-\theta)$ is implemented depends on the previous measurement outcome $s_1$ and in general all previous measurement outcomes for an arbitrary long sequence of gates. Since we can determine  beforehand if there is a flip we will measure the observable $A(-\theta)$ instead of $A(\theta)$ to compensate for it. The general single qubit unitary in Euler normal form $U(\zeta, \eta, \xi)$, where $U(\zeta, \eta, \xi) = e^{-\zeta X / 2}e^{-\eta Z / 2}e^{-\xi X / 2}$, is implemented using a 5 qubit linear cluster as shown in figure~\ref{fig:sqr_cluster}. CNOT gates can be implemented using a 15 qubit cluster state as shown in figure~\ref{fig:cnot_cluster}.

Finally, the preparation of the state $\ket \psi$ itself can be incorporated into the protocol by starting with a suitable single qubit rotation. Since arbitrary single qubit gates and CNOT gates together form a universal set of gates, any quantum circuit can be simulated on a 2D cluster state. 

\subsection{Circuit transpilation}\label{sec:circ_transpilation}

A high level circuit, described in terms of gates acting on qubits, can be implemented using a 2D square lattice cluster state of sufficient size. In the following, we describe the implementation method.

\begin{figure}
    \centering
    \includegraphics[width=\linewidth]{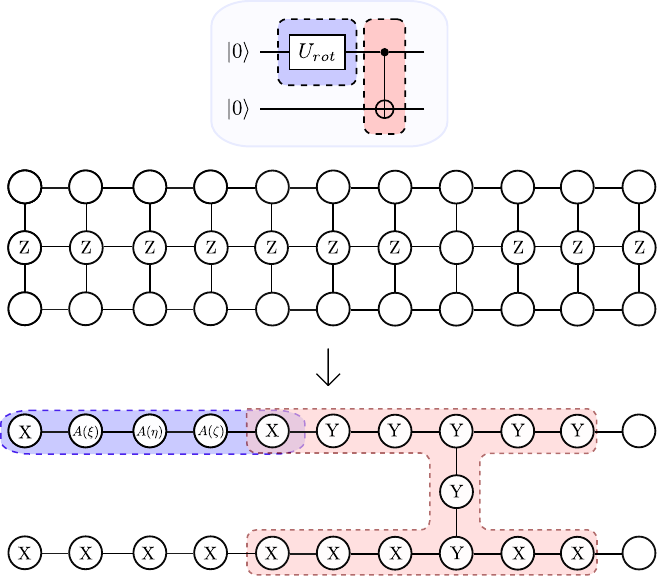}
    \caption{The logical circuit is transpiled for MBQC by 1) starting with a lattice cluster state of sufficient size. 2) Performing Pauli Z measurements to disconnect qubits and produce the required topology. 3) All Pauli measurements are performed in the first round of measurement (implements all Clifford gates). 4) Adaptive measurements are performed on the remaining qubits over several rounds until the computation is finished. 5) Post processing of the measurement results recovers the output of the logical circuit. }
    \label{fig:cluster_etching}
\end{figure}

Assuming that a circuit consists of single qubit rotations and nearest neighbour CNOTs (which any circuit can be implemented as) the first step is to etch out the circuit topology out of the lattice cluster state. This is done using Pauli Z measurements which disconnect the measured node from the cluster as shown in figure~\ref{fig:cluster_etching}.  
Once we have this topology which will support the gates what remains is to measure the remaining qubits in an adaptive fashion to simulate the gate operations. 

One could in principle implement the gates in the same order as they are done in the circuit description. However, it turns out this is not necessary, or indeed even preferable \cite{raussendorf_computational_2002}. The basis in which to measure a particular qubit $k$ depends on the measurement outcomes of a set of qubits $b_k$ which is called the backward cone of the qubit $k$. This implies that the  qubits in the cluster state can be partitioned into disjoint sets $\{Q_t\}$ determining different measurement rounds. The measurement of the set $Q_0$ which contains qubits who's measurement basis does not depend on the outcome of any other measurement is conducted first. The results of this measurement round is used to determine the basis to measure qubits in $Q_1$, and so on until the computation is finished. 

The set $Q_0$ contains all the qubits on which Pauli measurements are to be performed. The reason for this is that the adaptive nature of the measurement basis is governed by the presence of a Pauli operator $\sigma_i \in \{X, Y, Z \}$ on the qubit to be measured. The measured observable $\sigma_j$ is transformed by this Pauli operator as $\sigma_i \sigma_j \sigma_i = (-1)^{\delta_{ij}} \sigma_j$. In other words, we either measure $\sigma_j$ or $-\sigma_j$ which is just a relabelling of the measurement outcomes. All Clifford operations can be implemented in the MBQC scheme with just Pauli measurements, which means that the first round of measurement $Q_0$ implements all Clifford gates in the circuit. Note that this does not depend on the temporal ordering of these gates in the actual quantum circuit; in fact even the final computational basis state measurements at the end of the circuit are implemented in this first round. This doesn't mean that the output of the quantum circuit is known after the first round, the measurement outcomes of all the qubits are required to reconstruct the output of the logical circuit. As a consequence of this measurement based implementation of the Clifford gates, there will be Pauli corrections scattered throughout the circuit which have to be propagated either to the end of the circuit to modify the the final measurement outcomes or to the start of the circuit to initialise the Pauli frame which tracks the accumulation of Pauli operators on the qubits.  

Since the set of stabilizer states is invariant under Pauli measurements, after the first round of measurement we are left with a stabilizer state $\ket{\psi_S}$ which is local Clifford equivalent to some graph state(s) \cite{van_den_nest_graphical_2004, schlingemann_stabilizer_2001}. These Clifford equivalent set of graph states are referred to as algorithm-specific graph states and can be considered the fundamental resource required to implement the quantum logic circuit.

\subsection{Algorithm based graph compilation}
Instead of preparing the algorithm-specific graph state through measurement as described in section \ref{sec:circ_transpilation}, we could instead efficiently compute it classically and prepare it directly. This is a consequence of the fact that all the operations that generate the algorithm-specific graph state are Clifford operations on stabilizer states and the Gottesman-Knill theorem \cite{gottesman_heisenberg_1998}. In this manner we are not performing any classically simulatable operations on a quantum device but saving those resources for the genuinely quantum part.

The most obvious way to generate an algorithm-specific graph is to start with a stabilizer description of a cluster state and following the procedure shown in figure~\ref{fig:cluster_etching} etch out the required circuit topology with $Z$ basis measurements and perform all the Pauli basis measurements implementing the measurement round $Q_0$. The obtained stabilizer state can be converted to a graph state using the procedure explained in section \ref{sec:graph_conversion}.

Alternatively, we can go one step further and not encode the Clifford gates into the cluster at all but simulate them as unitary gates at the logic circuit level before we encode the circuit into a graph state. To do this we recast the circuit in a form called the ICM (Initialise-CNOT-Measurement) form as described in section \ref{sec:ICM} which allows all Clifford gates to be implemented first at the logical circuit level. Our method is described in Section~\ref{sec:algo_examples}.

\section{Stabilizer state to Graph state conversion}
\label{sec:graph_conversion}

Every stabilizer state can be converted to a graph state using local Clifford operations. This means that given any stabilizer tableau, we can perform a process akin to Gaussian elimination augmented with local Clifford operations to obtain a corresponding graph state \cite{van_den_nest_graphical_2004}. Note that this graph state is not unique and there could be multiple graph states which \madhav{are local Clifford equivalent to a given stabilizer state.}

\subsection{Graph conversion algorithm}
\label{sec:graph_conv}

We now present an algorithm to convert any stabilizer state into a graph state (a pseudo-code is given in appendix \ref{apx:graph_conversion}). The algorithm uses the fact that modulo 2 addition of the rows of the tableau corresponds to multiplication of the corresponding stabilizers, up to phase correction. i.e. for a stabilizer set $ S = \{s_1, s_2,\dots,s_n \}$, we can replace the row corresponding to $s_j$ in the tableau with $rowsum(i, j)$, where $rowsum$ is a function that does modulo 2 addition for the $X$ and $Z$ blocks of the two rows and computes the appropriate phase for the final column. An efficient way to compute this phase from the tableau is given in \cite{aaronson_improved_2004}. To convert an arbitrary stabilizer to a graph state, we need to transform the $X$ block into an identity matrix, and the $Z$ block will automatically be symmetric to ensure that all the stabilizers commute \cite{van_den_nest_graphical_2004}. The full algorithm is as follows  ---

\vspace{0.3cm}

\noindent \textbf{Remove zero columns for the X block}:
\begin{enumerate}[label=\alph*)]

\item  If there is a zero column in the $X$-block, we perform a Hadamard operation by swapping that $X$ column with its corresponding $Z$ column since this $Z$ column has to be non-zero. This is because a valid stabilizer state set cannot have both the $Z$ and $X$ column be all zeros for a qubit. That would imply that there is only an identity stabilising that qubit for all the stabilizers in the set which would leave the state of the qubit unspecified.
\end{enumerate}
\textbf{Make the X block lower triangular}:
\begin{enumerate}[label=\alph*)]
\setcounter{enumi}{1}
\item Iterating $i$ from 1 to $n$ , if $X_{ii}$\footnote{Here the first subscript refers to the row and second the column.}  = 1, we can set all $X_{ji}$ with $j > i$ to equal 0 by replacing all rows $j$ such that $X_{ji} = 1$ with $rowsum(i,j)$. We do this for all rows  $j > i$. 

\item If for some $i$, $X_{ii} = 0$, but there exists some $j > i$ such that $X_{ji} = 1$, we swap rows $i$ and $j$ and perform the logic in step b).

\item If there is no $j \geq i$ such that $X_{ji} = 1$, there must exist some $Z_{ji} = 1$, since we know that for every stabilizer state the $X$ block can be made full rank using local operations and $rowsum$s, the only freedom left is to bring it from the $Z$ block. We perform a Hadamard operation swapping the $i^\mathrm{th}$ column of the $X$ and $Z$ block to achieve this and perform the logic in step b) or c) depending on whether $X_{ii}$ is 0 or 1. 

\end{enumerate}

\noindent \textbf{Make the X block diagonal}:
\begin{enumerate}[label=\alph*)]
  \setcounter{enumi}{4}
    \item For $i$ from $n$ to 1, if any $X_{ji} = 1$, replace row $j$ with $rowsum(i,j)$.
\end{enumerate}

\noindent\textbf{ Make Z block diagonal zero}:
\begin{enumerate}[label=\alph*)]
  \setcounter{enumi}{5}
    \item For all $i$ such that $Z_{ii} = 1$, we perform the phase gate transformation on qubit $i$ which performs the transformation $XZ \rightarrow X$, leaving us with $Z_{ii} = 0$.
\end{enumerate}

\noindent \textbf{Make all phases positive}:
\begin{enumerate}[label=\alph*)]
  \setcounter{enumi}{6}
    \item For all rows with a negative phase we apply a Z gate transformation which takes $X \rightarrow - X$.
\end{enumerate}

\subsection{Clifford circuit graph evolution example: GHZ}

If we start with a Clifford circuit, we can apply the conversion algorithm directly. For example, taking the three qubit GHZ generation circuit, the steps for converting the GHZ state to a graph state are shown in Tab.~\ref{tb:ghz_conversion}. The GHZ circuit,
\[
\begin{tikzpicture}
\node[scale=1]{
\begin{quantikz}
\lstick{$\ket{0}$}   & \gate{H} &\ctrl{1} &\qw &\qw \\
\lstick{$\ket{0}$}  & \qw &\targ{} & \ctrl{1} &\qw \\
\lstick{$\ket{0}$}  &\qw &\qw &\targ{} &\qw
\end{quantikz}
};
\end{tikzpicture}
\]
prepares the state $\ket{\psi} = \ket{000} + \ket{111}$ (we assume implicit normalisation throughout this paper for simplicity). Starting from the canonical stabilizer for the all zero state, the circuit does the following stabilizer transformation,
\[
\setlength\arraycolsep{0mm}\renewcommand\arraystretch{0.75}
\begin{matrix}
Z &I &I \\
I &Z &I \\
I &I &Z
\end{matrix} \rightarrow  \begin{matrix}
X &X &X \\
Z &Z &I \\
I &Z &Z
\end{matrix}.
\]
  
\begin{table}
    \centering
\begin{tabular}{|c|c|c|} \hline
    Operation & Stabilizer & State  \\ \hline
    & \setlength\arraycolsep{0mm}\renewcommand\arraystretch{0.75} $\begin{matrix}
              Z &I &I \\
              I  &Z &I \\
              I &I &Z \\
               \end{matrix}$ & $\ket{000}$  \\ \hline
$\begin{tikzcd}
\lstick{1}   & \gate{H} &\ctrl{1} &\qw &\qw \\
\lstick{2}  & \qw &\targ{} & \ctrl{1} &\qw \\
\lstick{3}  &\qw &\qw &\targ{} &\qw
\end{tikzcd}$ & \setlength\arraycolsep{0mm}\renewcommand\arraystretch{0.75}$\begin{matrix}
             X &X &X \\
             Z &Z &I \\
             I &Z &Z \\
               \end{matrix}$ &  $\ket{000} + \ket{111}$       \\ \hline
               $H$ on 2 & \setlength\arraycolsep{0mm}\renewcommand\arraystretch{0.75} $\begin{matrix}
              X &Z &X \\
              Z &X &I \\
              I &X &Z \\
               \end{matrix}$ & $\ket{0+0} + \ket{1-1}$  \\ \hline
              $rowsum(2\rightarrow 3)$ & \setlength\arraycolsep{0mm}\renewcommand\arraystretch{0.75} $\begin{matrix}
              X &Z &X \\
              Z &X &I \\
              Z &I &Z \\
               \end{matrix}$ & $\ket{0+0} + \ket{1-1}$ \\ \hline
               $H$ on 3 & \setlength\arraycolsep{0mm}\renewcommand\arraystretch{0.75} $\begin{matrix}
              X &Z &Z \\
              Z &X &I \\
              Z &I &X \\
               \end{matrix}$  & $\ket{0++} + \ket{1--}$  \\ \hhline{|= = =|}
              \multicolumn{1}{|c|}{Final Graph} &  \multicolumn{2}{c|}{ \parbox[c]{3cm}{\includegraphics[width=\linewidth]{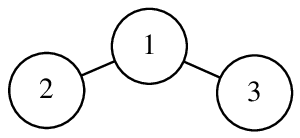}}} \\ \hline
\end{tabular}
\caption{Conversion of a 3 qubit GHZ state to a graph state. The action of a Hadamard on a qubit is to swap $Z$s and $X$s for the column corresponding to the qubit. $rowsum(i\rightarrow j)$ replaces $s_j$ with $s_is_j$. }
\label{tb:ghz_conversion}
\end{table}
\madhav{as shown in the first two rows of Tab.~\ref{tb:ghz_conversion}. The remaining rows of Tab.~\ref{tb:ghz_conversion} describes the conversion of this stabilizer state to a graph state through the application of Clifford gates (\textit{rowsums} do not correspond to physical operations). Since this conversion process is invertible we can instead view this in the opposite direction --- beginning with the algorithm-specific graph state in the final row of Tab.~\ref{tb:ghz_conversion} and applying the inverse of the local transformation rules (in this case Hadamard gates on qubits 3 and 2) generates the output of the Clifford circuit.}

\section{A compiler for algorithm-specific graph states}  
\label{sec:algo_examples}

We introduce a circuit compiler formalism which allows us to systematically decompose any given circuit using T-state injection \cite{bib:NielsenChuang00} and a variation of it using T gate teleportation which will form part of our compiler pipeline. 

In this section, we show that the ability to perform all Clifford operations simultaneously in the first round of measurement in the MBQC model has a corresponding analogue in the circuit model. Here, instead of initialising a standard cluster state and performing Pauli measurements we implement non-Clifford gates using gate teleportation as shown in figure~\ref{Fig:invicm-tgate}, which allows all Clifford gates to be implemented first.

We will also provide a step-by-step derivation of an algorithm-specific graph state for some typical quantum gates. The basic principles demonstrated can be extended to arbitrary quantum circuits. 

In the MBQC model, the Clifford gates could be implemented first because of the invariance of Pauli measurement bases due to Pauli corrections as explained in sec \ref{sec:circ_transpilation}. The corresponding property that allows this in the circuit model is that Pauli corrections due to the teleported non-Clifford gates when commuted through a Clifford gate do not change the Clifford gate but only the correction. This allows us to track the corrections and propagate them to the end of the circuit directly before subsequent measurements. This will impose a temporal ordering on the measurements that implement the teleported gates since the Pauli corrections due to previous measurements will determine the measurement bases for current measurements.

\subsection{ICM formalism}
\label{sec:ICM}
The ICM form of a quantum circuit (figure~\ref{Fig:icm-formalism}) was \madhav{ introduced in \cite{paler_fault-tolerant_2017}  for  logical level QEC compilation.} It is a decomposition of an arbitrary circuit into three layers --- 1) an initialisation layer preparing all qubits into one of 4 states, 2) an array of CNOTs and 3) a staggered measurement block in the $Z$ and $X$ basis with the basis choice of subsequent blocks depending on the measurement outcomes of previous ones. 
\begin{figure}
\begin{tikzpicture}
\node[scale=1]{
\begin{quantikz}
\lstick{} &\gate[wires=4]{I} & \gate[wires=4]{CNOT}& \gate[wires=2]{M_{X/Z}} &  \\
  & & & \qw & \cw \cwbend{1} \\
\lstick{}  & &   & \qw &\gate[2]{M_{X/Z}} \\
\lstick{}  & & &\qw & \qw &\cw   
\end{quantikz}
};
\end{tikzpicture}
\caption{ICM form of a quantum circuit. The $I$ block initialises each qubit in either the state
$\ket{0}$ or the states $\ket{0} + e^{i k \pi }\ket{1}$, with $k \in \{ 0, 1/2, 1/4 \}$. The CNOT block is an array of CNOTs. The measurement blocks performs measurement in the $X$ or $Z$ basis and subsequent measurements bases are decided based on the outcome.}
\label{Fig:icm-formalism}
\end{figure}
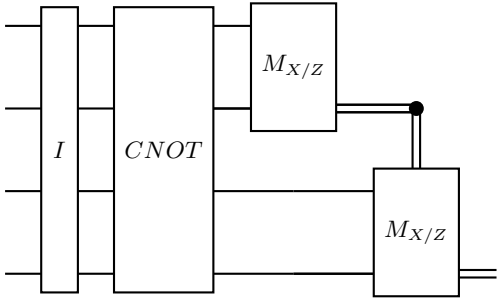
\begin{figure}
\begin{tikzpicture}
\node[scale=1.2]{
\begin{quantikz}
\lstick{$\ket{\psi}$}   & \targ{} &\meter{$Z$} \arrow[r] &\rstick{$s$} \\
\lstick{$\ket{A}$}  & \ctrl{-1} &\qw & \qw \rstick{$X^s P^s T\ket{\psi}$} 
\end{quantikz}
};
\end{tikzpicture}
\caption{T gate implementation using state injection in the ICM formalism. Here the injected state $\ket A = T\ket{+}$. The X correction is propagated forward to measurement while the $P$ gate correction is implemented by a similar state injection of $\ket Y = P\ket +$ instead of $\ket{A}$.}
\label{Fig:icm-tgate} 
\end{figure}
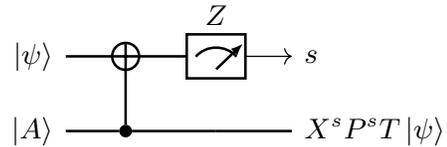

The ICM formalism can be understood to be systematic state injection with selective measurement corrections applied through the basis choice in the measurement blocks.  While the state injection circuit in figure~\ref{Fig:icm-tgate} can be verified directly, it can also be understood from the basic MBQC gate teleportation circuit in figure~\ref{fig:gate_teleportation}. Using the fact that $\ket A = T \ket +$ and that $Z$ measurement is equivalent to a Hadamard gate followed by an $X$ measurement, the circuit in figure~\ref{Fig:icm-tgate} can be redrawn as,

\[
\begin{tikzpicture}
\node[scale=1]{
\begin{quantikz}
\lstick{$H\ket{\psi}$} & \gate{H}       & \targ{}    &\gate{H} &\meter{$X$} \arrow[r] &\rstick{$s$} \\
\lstick{$\ket{+}$}     & \qw            & \ctrl{-1}  & \qw &\gate{T} & \qw  
\end{quantikz}.
};
\end{tikzpicture} 
\]
Here we have commuted the $T$ gate through the control of the CNOT gate. Using the fact that conjugating the target of a CNOT gate with a Hadamard induces a $CZ$ gate, we can see this circuit is equivalent to the gate teleportation circuit in figure~\ref{fig:gate_teleportation}  with an input state $H\ket\psi$, $U_Z(\theta) = \mathbb I$ \madhav{(i.e., $A(\theta = 0) = X)$} and a $T$ gate acting on the output. Hence, the output of this circuit must be $THZ^sH\ket{\psi} = TX^s\ket\psi $. When the measurement outcome is 0, we have a $T$ gate applied as desired but when the outcome is $1$ there is an $X$ operator acting before T which we need to commute to the outside using the equivalence (up to global phase) $TX = XT^{\dagger}$. This gate can be corrected to a $T$ gate by applying the operator $PX$ hence confirming the circuit output in figure~\ref{Fig:icm-tgate}.   

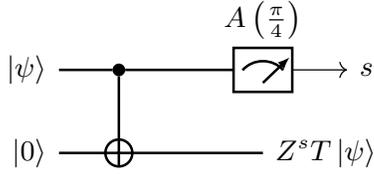
\begin{figure}[h]
\begin{tikzpicture}
\node[scale=1.2]{
\begin{quantikz}
\lstick{$\ket{\psi}$}   & \ctrl{1}  &\qw    &\meter{$A\left(\frac{\pi}{4}\right)$}  \arrow[r] &\rstick{$s$} \\
\lstick{$\ket{0}$}      & \targ{}   &\qw    &\qw  \rstick{$Z^sT\ket{\psi}$} 
\end{quantikz}
};
\end{tikzpicture}
\caption{$T$ gate teleportation in the inverse ICM formalism where the measured observable is $A\left(\frac{\pi}{4}\right) = T^{\dagger}XT$. $T^{\dagger}$ can be similarly implemented by measuring the observable $A\left(-\frac{\pi}{4}\right) = TXT^{\dagger}$. Using this subcircuit we are able to decompose any Clifford + T circuit into a stabilizer state initialisation followed by non-stablizer measurements.}
\label{Fig:invicm-tgate}
\end{figure}

The CNOT and measurement blocks of the ICM form are Clifford operations and Pauli measurements, but the initialisation block requires the preparation of non-Clifford states for state injection. For our purpose we modify this scheme as shown in \cite{herr_lattice_2017} known as inverse ICM, such that the initialisation block only prepares stabilizer states, but as a trade-off the measurements are now non-Pauli measurements. This is achieved using the gate teleportation circuit shown in figure~\ref{Fig:invicm-tgate}. This circuit can be derived from the gate teleportation circuit in figure~\ref{fig:gate_teleportation} by absorbing the Hadamard gate from the output to the right of the $CZ$ gate and the Hadamard from the input $\ket +$ state to the left of it and setting $U_Z(\theta) = T$.

\subsection{Controlled-$V^{\dagger}$ graph states}
\begin{figure}
\begin{tikzpicture}
\node[scale=1]{
\begin{quantikz}[column sep=0.351cm]
\lstick{$\ket{a}$}   & \ctrl{1} &\qw \\
\lstick{$\ket{b}$}  & \gate{
V^{\dagger}} &\qw 
\end{quantikz} = 
\begin{quantikz}[column sep=0.251cm]
\lstick{$\ket{a}$}   &\gate{T^{\dagger}} &\targ{}     &\gate{T}           &\targ{}     &\qw  &\qw\\
\lstick{$\ket{b}$}  &\gate{H}            &\ctrl{-1} &\gate{T^{\dagger}}  &\ctrl{-1} &\gate{H} &\qw 
\end{quantikz} 
};
\end{tikzpicture}
\caption{Controlled-$V^{\dagger}$ decomposition using Clifford + T gates.}
\label{Fig:control-V}
\end{figure}
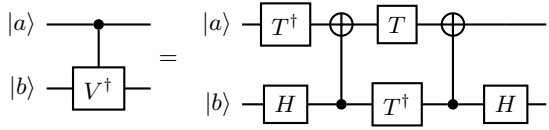

Consider the Clifford + $T$ gate decomposition of the controlled-$V^\dagger$ gate show in figure~\ref{Fig:control-V}. The $V^{\dagger}$ gate is one of the square roots of the X gate and is defined as,
\begin{align}
    V^{\dagger} = \frac{1}{2}\begin{pmatrix}
    1-i & 1 + i \\
    1+i & 1 - i
    \end{pmatrix}.
    \label{eq:Vdag_def}
\end{align}

The $T$ and $T^{\dagger}$ gates are implemented using the teleportation algorithm in figure~\ref{Fig:invicm-tgate} with the Pauli corrections propagated to the end of the circuit, as shown in figure~\ref{Fig:inverse_icm_example}. {\color{black} 
\begin{figure}
\begin{tikzpicture}
\node[scale=0.9]{
\begin{quantikz}[column sep=0.251cm]
\lstick{$\ket{a}$}  &\ctrl{2}   &\qw  &\qw      &\qw    &\qw     &\qw   &\qw  &\qw   &\qw &\qw  &\meter{$A\left(-\tfrac{\pi}{4}\right)$ } \arrow[r] &\rstick{$r$}\\
\lstick{$\ket{b}$}  &\qw  &\gate{H}  &\ctrl{1}    & \ctrl{3}  &\qw  &\qw &\qw &\qw &\qw &\gate{Z^r}   &\meter{$A\left(-\tfrac{\pi}{4}\right)$ } \arrow[r] &\rstick{$s$}\\
\lstick{$\ket{0}$}  &\targ{}    &\qw    &\targ{}        &\qw    &\qw       &\qw &\ctrl{1}     &\qw &\qw  &\gate{Z^r}  &\meter{$A\left(\tfrac{\pi}{4}\right)$ } \arrow[r] &\rstick{$t$} \\
\lstick{$\ket{0}$}  &\qw   &\qw     &\qw &\qw  &\qw  &\qw  &\targ{} &\targ{} &\qw &\gate{Z^s}  &\qw \rstick{$a_{out}$}\\
\lstick{$\ket{0}$}  &\qw        &\qw            &\qw        &\targ{}    &\qw     &\qw &\qw  &\ctrl{-1}      &\gate{H}  & \gate{X^{t\oplus s}}  &\qw \rstick{$b_{out}$}
\end{quantikz} 
};
\end{tikzpicture}
\caption{Teleported $T$ gate implementation of the controlled-$V^{\dagger}$ gate with the Pauli corrections propagated to the end.}
\label{Fig:inverse_icm_example}
\end{figure}
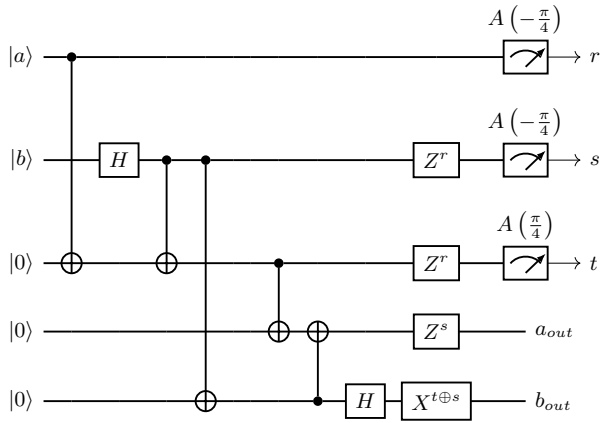
In this example, all the non-stabilizer measurements can be performed simultaneously since the only Pauli corrections for the measured qubits are $Z$ corrections that changes the measurement of $A(\theta) \rightarrow ZA(\theta)Z = - A(\theta)$, which is a classical relabelling of the measurement outcomes. However, this is not the case in general, for example, if we wanted to now implement a $T$ gate on the qubit $b_{out}$, the presence or absence of the X correction will decide the measurement basis since $X A(\theta) X = A(-\theta)$. Note that in the state injection picture of the ICM form, measurement outcomes determined whether $T$ gates or $T^{\dagger}$ gates were implemented due to the Pauli $X$ correction, we can see the same holds true here. We will not apply unitary corrections for these Pauli gates but rather modify the the measurement basis to account for them. For example if we know based on previous measurement that an $X$ gate correction acts on a qubit before an $A(\theta)$ basis measurement, we will instead perform an $A(-\theta)$ basis measurement. These Pauli corrections thus give a time ordered measurement pattern with feed-forward which can be determined at compilation time using Pauli tracking \cite{paler_software-based_2014}.
}

\begin{figure}
    \centering
\includegraphics[width=0.65\linewidth]{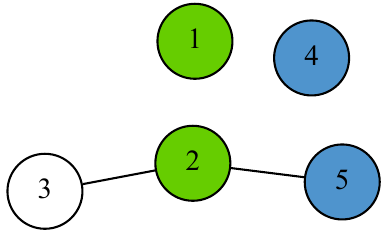} 
    \caption{Local Clifford equivalent graph state before measurement for the controlled-$V^{\dagger}$ decomposition in figure~\ref{Fig:inverse_icm_example} for inputs $\ket{ab} = \ket{00}$. The input\footnote{Note that input here refers to the labeling of the circuit qubits; the input state is not encoded into the green nodes, rather different inputs lead to different graph structures. } and output qubits are shown as green and blue, respectively. The circuit is implemented by applying local Clifford gates and measuring all the qubits except the output qubits.}
    \label{fig:ctrl_v_graph}
\end{figure}

For inputs $\ket{ab} = \ket{00}$, the state of the system before the measurements are performed, ignoring the Pauli corrections is $\ket{0000+} + \ket{0110-}$, where $\ket{\pm} = \ket{0} \pm \ket{1}$. This is locally equivalent to the graph state shown in figure~\ref{fig:ctrl_v_graph} (see App.~\ref{apdx:ctrl_v_example} for a detailed derivation).
 We can equivalently start with the graph state and apply the local corrections to generate the same output using the circuit given in figure~\ref{Fig:ctrl-v_compiled}.
 {\color{black}
 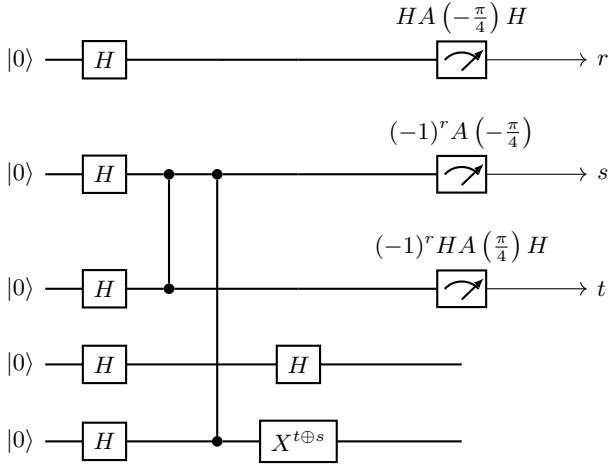
\begin{figure}
\begin{tikzpicture}
\node[scale=1]{
\begin{quantikz}
\lstick{$\ket{0}$} & \gate{H}   &\qw         &\qw        & \qw  &\meter{$HA\left(-\tfrac{\pi}{4}\right)H$ } \arrow[r] &\rstick{$r$}\\
\lstick{$\ket{0}$} & \gate{H}  &\ctrl{1}    &\ctrl{3}        &\qw   &\meter{$(-1)^rA\left(-\tfrac{\pi}{4}\right)$ } \arrow[r] &\rstick{$s$}\\
\lstick{$\ket{0}$} & \gate{H}  &\control{}        &\qw        &\qw &\meter{$(-1)^rHA\left(\tfrac{\pi}{4}\right)H$ } \arrow[r] &\rstick{$t$}  \\
\lstick{$\ket{0}$} & \gate{H}  &\qw  & \qw  &\gate{H}   &\qw \\
\lstick{$\ket{0}$} & \gate{H}  &\qw         &\control{} &\gate{X^{t\oplus s}}        &\qw\\ 
\end{quantikz}
};
\end{tikzpicture}
\caption{Compiled circuit: The Hadamard and the CZ gates prepare the graph state given in figure~\ref{fig:ctrl_v_graph} and then the local corrections are applied to generate the same output as the circuit in figure~\ref{Fig:inverse_icm_example}. In practise the local corrections can be absorbed into the measurement basis choice as shown for the first 3 qubits. The remaining corrections are also similarly propagated forward until measurement.}
\label{Fig:ctrl-v_compiled}
\end{figure}
}
The intuition from the graph in figure~\ref{fig:ctrl_v_graph} is that qubits 1 and 4 should factor out in the pre-measurement state -- indeed, this is true for the state $\ket{0000+} + \ket{0110-}$.

To perform the controlled-$V^{\dagger}$ gate we implement the measurements in the $A\left(-\tfrac{\pi}{4}\right)$ basis on qubits 1, 2 and $A\left(\tfrac{\pi}{4}\right)$ on qubit 3. For ease of calculation, we apply the local corrections explicitly rather than absorbing them into the measurement basis and also assume the measurement outcomes $r=t=s = 0$. This can always be achieved in practice by measuring the qubits 2 and 3 in the basis of $-A\left(-\tfrac{\pi}{4}\right)$  and $-A\left(\tfrac{\pi}{4}\right)$, respectively, if $r=1$ and applying a $X$ correction to qubit 5 if $t\oplus s = 1$.

The operators $A\left(\pm\tfrac{\pi}{4}\right)$ can be written in their spectral decomposition as,
\begin{align}
    A\left(\tfrac{\pi}{4}\right)  &= \op{A^*}{A^*} - \op{\ov A^*}{\ov A^*}, \\
    A\left(-\tfrac{\pi}{4}\right) &= \op{A}{A} - \op{\ov A}{\ov A}
\end{align}
where, $\ket{A} = T\ket+$, $\ket{\ov A} = T\ket-$ and $\ket{ A^*}$, $\ket{ \ov A^*}$ are similarly defined with $T^{\dagger}$ instead of $T$. Measuring $A\left(\pm\tfrac{\pi}{4}\right)$ as 0, projects the computational basis states as,
\begin{align}
    \ket{A^*} \braket{A^* |  0} &= \ket{A^*} , &\ket{A^*} \braket{A^* | 1} &= \ e^{i\pi/4}\ket{A^*}, \\
    \ket{A} \braket{A |  0} &= \ket{A} , &\ket{A} \braket{A | 1} &= \ e^{-i\pi/4}\ket{A}.
\end{align}
 The measurement projects the state as,
 \begin{align}
     & \op{A}{A} \otimes \op{A}{A} \otimes \op{A^*}{A^*}\left( \ket{0000+} + \ket{0110-} \right) \notag\\ 
     =&\ket{AAA^*} \left( \ket{0+} + \ket{0-} \right) = \ket{AAA^*00}.
 \end{align}
giving the output state in the last two qubits as, 
\begin{align}
\ket{a_{out}b_{out}} =  \ket{00}.
\end{align}
This is of course the output of a controlled-$V^\dagger$ acting on the input $\ket{ab} = \ket{00}$. A less trivial example is given by choosing the input to be $\ket{ab} = \ket{+0}$. This is equivalent to applying a Hadamard gate to the first qubit at the beginning of the circuit in figure~\ref{Fig:inverse_icm_example}. Since this only initialises a different stabilizer state and all Pauli corrections are introduced after this point, the measurement sequence and how future measurements are influenced by previous \madhav{measurement outcomes} remain unchanged. In this case, the stabilizer state produced before measurement is given by,
\begin{align}
    \ket{\psi} = \ket{0000+} + \ket{0110-} +\ket{1011+} + \ket{1101-}
\end{align}
which can be converted to the graph state shown in figure~\ref{fig:ctrl-V_graph_+0} (see appendix \ref{apdx:ctrl_v_example} for a full derivation). 
\begin{figure}
    \centering
\includegraphics[width = 0.65\linewidth]{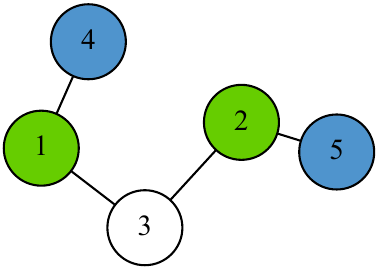}
    \caption{Locally equivalent graph state before measurement produced by the circuit in figure~\ref{Fig:inverse_icm_example} for inputs $\ket{ab} = \ket{+0}$. Input and output nodes are shown in green and blue, respectively.}
    \label{fig:ctrl-V_graph_+0}
\end{figure}
The post measurement state in this case will be, 
\begin{align*}
&\op{A}{A} \otimes \op{A}{A} \otimes \op{A^*}{A^*} \ket \psi \\
&=  \ket{AAA^*} \left( \ket{0+} + \ket{0-} +\ket{1+} + e^{-i\pi/2}\ket{1-} \right) \\
&= \ket{AAA^*} \left( \ket{00}  + (1 -i)\ket{10}  + (1+i)\ket{11} \right) 
\end{align*}
Normalising the state above and tracing out the measured qubits, we get the output state as,
\begin{align}
    \ket{a_{out}b_{out}} = \frac{1}{\sqrt{2}} \ket{00} +  \frac{1 - i }{ 2\sqrt{2}} \ket{10} + +  \frac{1 + i }{ 2\sqrt{2}} \ket{11}
\end{align}

The $V^\dagger$ gate acts on the computational basis states as,
\begin{align*}
     \ket{0} \xrightarrow{V^\dagger} \frac{1 -i}{ 2}\ket{0} + \frac{1 + i}{ 2}\ket{1} \\
     \ket{1} \xrightarrow{V^\dagger} \frac{1 + i}{2}\ket{0} + \frac{1 - i}{2}\ket{1} \\
\end{align*}
The action of the controlled-$V^{\dagger}$ on the initial state $\ket{+0}$ can be seen to be,
\begin{align*}
    \ket{+0} &\xrightarrow{CV^{\dagger}} \frac{1}{\sqrt 2} \left( \ket{00} + \ket{1} \left(  \frac{1 -i}{2}\ket{0} + \frac{1 + i}{2}\ket{1} \right) \right), \\
    &= \frac{1}{\sqrt 2} \ket{00} + \frac{1 - i }{2\sqrt 2}\ket{10} + \frac{1+i}{2\sqrt 2}\ket{11}
\end{align*}
as expected.

\subsection{Toffoli gate graph states}

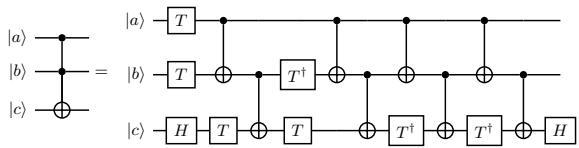
\begin{figure}
\begin{tikzpicture}
\node[scale=0.7]{
\begin{quantikz}[column sep=0.351cm]
\lstick{$\ket{a}$}  & \ctrl{2} &\qw \\
\lstick{$\ket{b}$}  & \ctrl{1} &\qw \\
\lstick{$\ket c$}   &\targ{}   &\qw
\end{quantikz} = 
\begin{quantikz}[column sep=0.251cm]
\lstick{$\ket{a}$}   &\gate{T}  &\ctrl{1} &\qw       &\qw                   &\ctrl{1}  &\qw         &\ctrl{1}           &\qw        &\ctrl{1}           &\qw        &\qw \\
\lstick{$\ket{b}$}   &\gate{T}  &\targ{}  &\ctrl{1}  &\gate{T^{\dagger}}    &\targ{H}  &\ctrl{1}    &\targ{}            &\ctrl{1}   &\targ{}            &\ctrl{1}   &\qw\\
\lstick{$\ket c$}    &\gate{H}  &\gate{T} &\targ{}   &\gate{T}              &\qw       &\targ{}     &\gate{T^{\dagger}} &\targ{}    &\gate{T^{\dagger}} &\targ{}    &\gate{H}
\end{quantikz} 
};
\end{tikzpicture}
\caption{Toffoli gate decomposition using Clifford + T gates.}
\label{fig:toffoli_decomp}
\end{figure}

We describe the decomposition of a Toffoli gate. We start with the seven T gate decomposition shown in figure~\ref{fig:toffoli_decomp}. We apply the same procedure as before and teleport the $T$  gates using the circuit in figure~\ref{Fig:invicm-tgate}. 
The graph states produced along with the local corrections and qubits to be measured are given in Tab.~\ref{tb:toffoli_graphs}. From the generated graphs it is clear that one can only generate entanglement between all the output qubits if both the controls of the Toffolli gate are active (not in the $\ket{0}$ state). 
\begin{table*}[]
    \centering
    \begin{tabular}{|c|c|c|} \hline 
       Input    & Graph & Local Corrections \\ \hline 
    $\ket{000}$ & \parbox[c]{5cm}{\includegraphics[width=\linewidth]{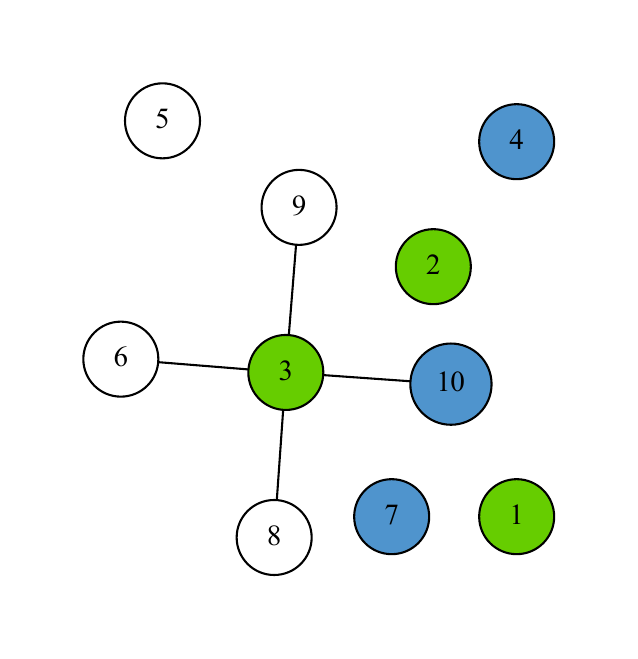}}      &    $H_1H_2H_4H_5H_6H_7H_8H_9$ \\ \hline
    $\ket{+00}$ & \parbox[c]{5cm}{\includegraphics[width=\linewidth]{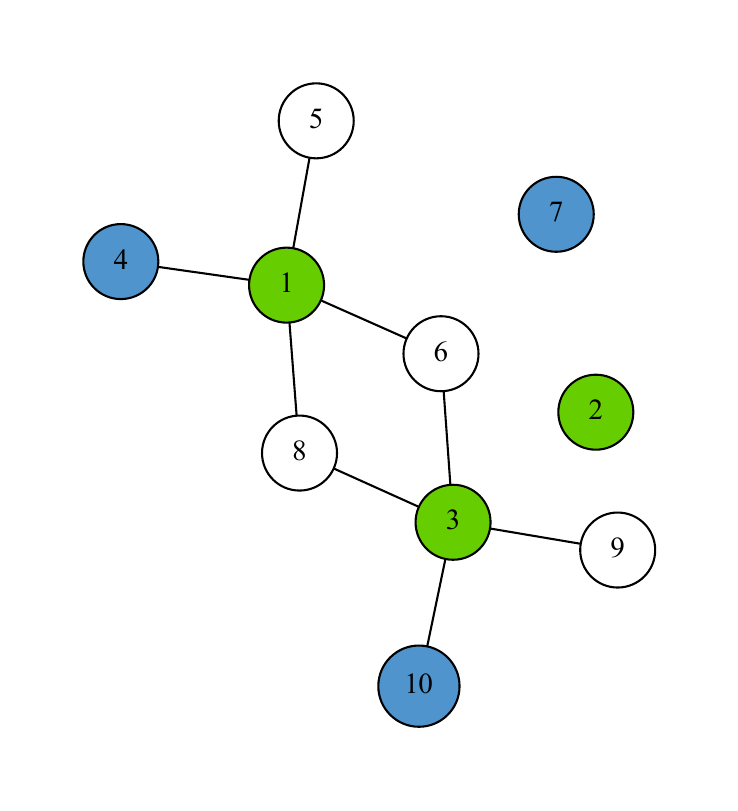}}      &    $H_2H_4H_5H_6H_7H_8H_9$ \\ \hline
    $\ket{++0}$ & \parbox[c]{5cm}{\includegraphics[width=\linewidth]{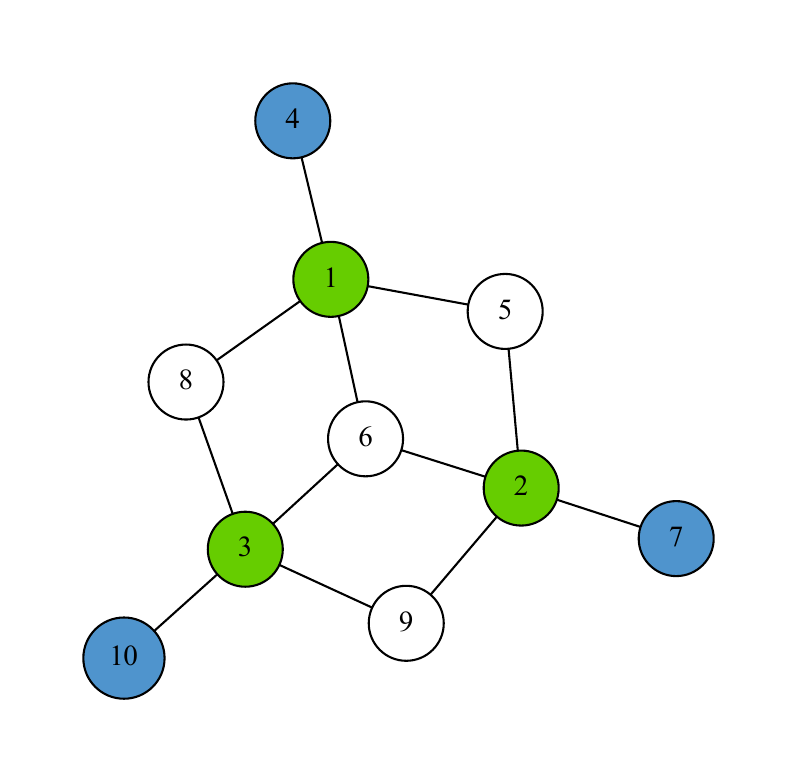}}      &    $H_4H_5H_6H_7H_8H_9$ \\ \hline
              \end{tabular}
    \caption{Toffoli gate implementation for different inputs. The input and output qubits are shown in green and blue respectively. The circuit is implemented by generating the graph state, applying the local corrections and measuring all the qubits except the output qubits in the $A\left(\pm\tfrac{\pi}{4}\right)$ basis. Pauli corrections to the measurements are assumed to be classically tracked through the circuit to determine the measurement sequence.   }
    \label{tb:toffoli_graphs}
\end{table*}

\section{Discussion}

It is important to note that the graph states produced by our algorithm are not unique. There can be many graph states that are equivalent under local Clifford operations \cite{van_den_nest_graphical_2004} and hence locally equivalent to the output stabilizer state of the circuit. This fact can be exploited to optimise over different graph state implementations. We now discuss this in more detail as well as consider the hardware compatibility of our approach.

\subsection{Resource optimisation}
\label{sec:resource-optimisation}
It has been shown that transformation between graph states using local Clifford operations is completely characterised by repeated application of the local complementation operations \cite{van_den_nest_graphical_2004}. In graph theoretic terms for a vertex $k$ with neighbourhood $n_k$, the local complementation will invert the edges (remove if there is one, add if there is not) between all vertices in $n_k$. Physically  this operation on vertex $k$ is performed by the operation $\sqrt{-iX_k}\prod\limits_{j \in n_k}\sqrt{iZ_j}$, where $\sqrt{-iX_k} = e^{-i\frac{\pi }{4}X_k} $ is the square root of $X$ operation acting on qubit $k$ and $\sqrt{iZ} = e^{i\frac{\pi }{4}Z_j}$ is the square root of $Z$ operation acting on qubit $j$. If two graph states $\ket{G}$ and $\ket{G^{\prime}}$ are equivalent under local Clifford transformation there exist some finite set of local complemention operations $\{l_i \}$ such that,
\begin{align}
    l_{n}\dots l_{2}l_{1}\ket{G} = \ket{G^{\prime}}.
\end{align}
Whether two graph states are indeed equivalent under local Clifford operations can be determined in polynomial time \cite{van_den_nest_efficient_2004}.

This gives us an additional degree of freedom to optimise our algorithm-specific graph state. Depending on the physical resources it might be desirable to reduce the total edge count, degree (number of edges per node) or prepare an initial state with a simple topology such as a linear graph state. All these problems are well studied in classical graph theory and find ready application here \cite{west2001introduction, adcock_mapping_2020, hoyer_resources_2006}.   

\subsection{Compatible architecture}\label{sec:compat_arch}

\madhav{MBQC based circuit execution is well suited to many optical quantum computing architectures \cite{bartolucci_fusion-based_2021} as well as NISQ devices \cite{ferguson_measurement-based_2021}. However, the compilation workflow presented here is not confined to any particular device architecture as one could use it as the logical level representation for any quantum device. An attractive application would be in the execution of fault tolerant quantum circuits using a suitable error corrected code such as the surface code \cite{fowler_surface_2012, horsman_surface_2012, bombin_topological_2010}. The nodes of the algorithm-specific graph will now be logical qubits in a protected subspace. This could offer several simplifications during run time as the only gates that need to be implemented are the ones that generate the graph state (H and CZ) and gates to implement fault tolerant rotated basis measurements dispensing with the need for lattice surgery (outside graph state generation) and dynamical ancillary routing operations \cite{litinski_game_2019}. This makes exact determination of the space-time volume of executing quantum algorithms in a fault tolerant manner much more tractable than it is presently.}

\section{Technical overview}
\label{sec:tech_overview}
In this section we describe the technical details of how different parts of the compiler pipeline are built. The ICM part of the pipeline has been built using Cirq, so Q\# circuits are first translated into Cirq circuits. Nevertheless, our compiler is output agnostic and can generate OpenQasm circuits, too.

\subsection{Q\# to Cirq conversion}
In order to support high-level Q\# algorithms \cite{heim_quantum_2020}, an intermediate component is needed to extract the circuit representation. Q\# is a functional language which is suited for hybrid quantum-classical algorithms and often exhibits non-trivial branching. 

The Q\# host program uses a simulator backend to perform compilation into a qubit register and a sequence of high-level operations. The choice of simulator backend depends on the nature of the experiment being run, but all are supported through the same translation extension. These operations are decomposed into primitive operations including \texttt{X, Y, Z, H, S, T, CX, R, Measure} gates. In the case of \madhav{ state space simulator backends}, operations are sometimes only decomposed into \texttt{SWAP, CCNOT, CCX, CCZ} gates.

To yield a circuit representation, an extension attaches a hook to the simulator before the experiment is run. During execution, a call is sent to the extension during each operation for parsing. If real-time operation is desired, the gates are broken down and streamed to the next stage of the pipeline. Otherwise, the circuit representation of the experiment from start to finish is stored in an output file. As our desired output format is Cirq, the stream of operations is converted into an initialisation of a Cirq qubit register and a sequence of moments. A standalone module implementing this part of the pipeline can be found at \url{https://github.com/QSI-BAQS/Trace2Cirq.git}.

\subsection{ICM compilation}
To perform an ICM decomposition, the compiler has to take a given Clifford + $T$ gate circuit and teleport all $T$ and $T^\dagger$ gates using ancillas, as described in Sec.~\ref{sec:ICM}. The Clifford + $T$ decomposition can be performed using Cirq's native decompose protocols and the main technical challenge left is when the compiler encounters a $T$ gate it needs to determine which wire this gate must act on depending on previous gate teleportations. For example, in the controlled-$V^\dagger$ decomposition in figure~\ref{Fig:inverse_icm_example}, when the compiler decomposes the $T$ gate, it needs to know that the decomposition is applied to the second wire if the $T^\dagger$ gate has already been decomposed. To do this the compiler adds an \texttt{op\_id} flag to every operation to keep track of the order of the operations and the class \texttt{SplitQubit} is used to track teleported qubits. When a gate is teleported, a new \texttt{SplitQubit} instance is created for the ancilla which inherits the \texttt{op\_id} of the operation that generated it and a reference to this new wire is stored in the original one. Comparing a gate's  \texttt{op\_id} with the those of the wire and it's children, the compiler is able to determine the correct wire to apply the decomposition on. 

\subsection{Stabilizer simulation}
The stabilizer simulator has been designed with a top level data structure which uses an integer array to track the tableau defined in Sec.~\ref{sec:tableau} and a backend which makes calls to a CHP simulator. This gives front-end access to the conceptually simpler tableau description while still using the more  efficient CHP formalism for actual computation. The stabilizer backend we use is STIM \cite{gidney_stim_2021}.

This part of the pipeline receives the inverse ICM decomposed Cirq circuit and computes the output stabilizer state produced by the circuit before the measurement sequence is applied. The graph conversion algorithm described in Sec.~\ref{sec:graph_conv} is now applied to the stabilizer state and the compiler returns the appropriate graph state and a list of local operations that convert it back to the output stabilizer state. The full compiler can be found here \url{https://github.com/QSI-BAQS/Jabalizer}.

\section{Conclusion}
\label{sec:conclusion}
We have presented a compiler to generate an algorithm-specific graph state based on a circuit description. It was shown that once the circuit is decomposed into a Clifford + $T$ gate circuit using standard quantum circuit libraries we can use a teleported $T$ gate implementation of the circuit. This allows the circuit to have a Clifford gate initialisation followed by non-Pauli measurements. The Clifford initialisation is simulated efficiently and an algorithm-specific graph state is extracted. Hardware level implementation of any circuit is now possible by generating this graph state, applying the local corrections and performing non-stabilizer measurements supplemented with tracking the Pauli frame. Our implementation is built for using Cirq circuits as input, but also accepts Q\# circuits which are internally converted to Cirq circuits. 

One of the main advantages of implementing the circuit in this way is that we are able to separate classical and quantum resources needed for a given quantum algorithm because the Clifford part of the circuit is classically simulated. The graph structure gives us an intuitive understanding of the entanglement structure that the algorithm utilised and how different qubits are connected and if some are superfluous. This leads naturally to questions of optimisation since the generated graph state is not unique. Depending on the physical hardware, different optimisations could be desirable such as reducing edge count, node degree etc. similarly to the co-design approach described in \cite{li_co-design_2021}.

It is an open question whether the algorithm-specific graph state generated using our method is the same as the the one generated by the traditional circuit etching method on the 2D cluster state, and is the subject of future work. 

\section*{Acknowledgements}

The views, opinions and/or findings expressed are those of the authors and should not be interpreted as representing the official views or policies of the Department of Defense or the U.S. Government.  This research was developed with funding from the Defense Advanced Research Projects Agency [under the Quantum Benchmarking (QB) program under award no. HR00112230007 and HR001121S0026 contracts].

\bibliography{refs}
%
%

\appendix

\section{Graph conversion of Controlled-$V^{\dagger}$}
\label{apdx:ctrl_v_example}
The evolution of the stabilizer state and the subsequent conversion to a graph state of the pre-measurement state of the circuit in figure~\ref{Fig:inverse_icm_example} is shown in Tabs.~\ref{tb:ctrl-v_clifford} and \ref{tb:ctrl-v_graph}.

\begin{table*}[!hbt]
\begin{minipage}[t]{.45\textwidth}
\resizebox{0.9\textwidth}{!}{%
\begin{tabular}[t]{|c|c|c|} \hline
\setlength\arraycolsep{0mm}
    Operation & Stabilizer & State  \\ \hline
                    & \setlength\arraycolsep{0mm} \renewcommand\arraystretch{0.75}  $\begin{matrix}
              &Z &I &I &I &I \\
              &I &Z &I &I &I \\
              &I &I &Z &I &I \\
              &I &I &I &Z &I \\
              &I &I &I &I &Z
               \end{matrix}$ &  $\ket{00000}$ \\ \hline
    $ \begin{tikzcd}
\lstick{1}   &  \ctrl{1} &\qw \\
\lstick{3}  & \targ{}  &\qw \\
\end{tikzcd}$ & \setlength\arraycolsep{0mm} \renewcommand\arraystretch{0.75} $\begin{matrix}
              &Z &I &I &I &I \\
              &I &Z &I &I &I \\
              &Z &I &Z &I &I \\
              &I &I &I &Z &I \\
              &I &I &I &I &Z
               \end{matrix}$ &  $\ket{00000}$ \\ \hline
    $ \begin{tikzcd}
\lstick{2}   &  \gate{H} &\qw \\
\end{tikzcd}$ & \setlength\arraycolsep{0mm} \renewcommand\arraystretch{0.75} $\begin{matrix}
              &Z &I &I &I &I \\
              &I &X &I &I &I \\
              &Z &I &Z &I &I \\
              &I &I &I &Z &I \\
              &I &I &I &I &Z
               \end{matrix}$ &  $\ket{00000} + \ket{01000}$ \\ \hline
     \begin{tikzcd}
\lstick{2}   &  \ctrl{1} &\qw \\
\lstick{3}  & \targ{}  &\qw \\
\end{tikzcd} & \setlength\arraycolsep{0mm} \renewcommand\arraystretch{0.75} $\begin{matrix}
              &Z &I &I &I &I \\
              &I &X &X &I &I \\
              &Z &Z &Z &I &I \\
              &I &I &I &Z &I \\
              &I &I &I &I &Z
               \end{matrix}$ &   $\ket{00000} + \ket{01100}$ \\ \hline
 \begin{tikzcd}
\lstick{3}   &  \ctrl{1} &\qw \\
\lstick{4}  & \targ{}  &\qw \\
\lstick{2}  &  \ctrl{1} &\qw \\
\lstick{5}  & \targ{}  &\qw \\
\end{tikzcd} & \setlength\arraycolsep{0mm} \renewcommand\arraystretch{0.75} $\begin{matrix}
              &Z &I &I &I &I \\
              &I &X &X &X &X \\
              &Z &Z &Z &I &I \\
              &I &I &Z &Z &I \\
              &I &Z &I &I &Z
               \end{matrix}$ &   $\ket{00000} + \ket{01111}$ \\ \hline
$ \begin{tikzcd}
\lstick{4}   &  \targ{} &\qw \\
\lstick{5}  & \ctrl{-1}  &\qw \\
\end{tikzcd}$ &  \setlength\arraycolsep{0mm}\renewcommand\arraystretch{0.75} $\begin{matrix}
              &Z &I &I &I &I \\
              &I &X &X &I &X \\
              &Z &Z &Z &I &I \\
              &I &I &Z &Z &Z \\
              &I &Z &I &I &Z
               \end{matrix}$ &   $\ket{00000} + \ket{01101}$  \\ \hline
$ \begin{tikzcd}
\lstick{5}   &  \gate{H} &\qw \\
\end{tikzcd}$ & \setlength\arraycolsep{0mm} \renewcommand\arraystretch{0.75} $\begin{matrix}
              &Z &I &I &I &I \\
              &I &X &X &I &Z \\
              &Z &Z &Z &I &I \\
              &I &I &Z &Z &X \\
              &I &Z &I &I &X
               \end{matrix}$ &   $\ket{0000+} + \ket{0110-}$ \\ \hline
\end{tabular}}
\caption{Gate transformations of the controlled-$V^{\dagger}$ decomposition give in figure~\ref{Fig:inverse_icm_example} excluding final measurements for inputs $\ket{ab} = \ket{00}$.}
\label{tb:ctrl-v_clifford}
\end{minipage}%
\begin{minipage}[t]{.45\textwidth}
\resizebox{0.9\textwidth}{!}{%
\begin{tabular}[t]{|c|c|c|} \hline
    Operation & Stabilizer & State  \\ \hline
                    & \setlength\arraycolsep{0mm} \renewcommand\arraystretch{0.75} $\begin{matrix}
              &Z &I &I &I &I \\
              &I &X &X &I &Z \\
              &Z &Z &Z &I &I \\
              &I &I &Z &Z &X \\
              &I &Z &I &I &X
               \end{matrix}$ &   $\ket{0000+} + \ket{0110-}$ \\ \hline
    $ \begin{tikzcd}
\lstick{1}   &  \gate{H} &\qw \\
\end{tikzcd}$ & \setlength\arraycolsep{0mm} \renewcommand\arraystretch{0.75} $\begin{matrix}
              &X &I &I &I &I \\
              &I &X &X &I &Z \\
              &X &Z &Z &I &I \\
              &I &I &Z &Z &X \\
              &I &Z &I &I &X
               \end{matrix}$ &   $\ket{+000+} + \ket{+110-}$ \\ \hline
$rowsum(1 \rightarrow 3)$ & \setlength\arraycolsep{0mm} \renewcommand\arraystretch{0.75} $\begin{matrix}
              &X &I &I &I &I \\
              &I &X &X &I &Z \\
              &I &Z &Z &I &I \\
              &I &I &Z &Z &X \\
              &I &Z &I &I &X
               \end{matrix}$ &   $\ket{+000+} + \ket{+110-}$  \\ \hline
    $ \begin{tikzcd}
\lstick{3}   &  \gate{H} &\qw \\
\end{tikzcd}$ & \setlength\arraycolsep{0mm} \renewcommand\arraystretch{0.75} $\begin{matrix}
              &X &I &I &I &I \\
              &I &X &Z &I &Z \\
              &I &Z &X &I &I \\
              &I &I &X &Z &X \\
              &I &Z &I &I &X
               \end{matrix}$ &   $\ket{+0+0+} + \ket{+1-0-}$  \\ \hline
$rowsum(3 \rightarrow 4)$ & \setlength\arraycolsep{0mm} \renewcommand\arraystretch{0.75} $\begin{matrix}
              &X &I &I &I &I \\
              &I &X &Z &I &Z \\
              &I &Z &X &I &I \\
              &I &Z &I &Z &X \\
              &I &Z &I &I &X
               \end{matrix}$ &   $\ket{+0+0+} + \ket{+1-0-}$  \\ \hline
$ \begin{tikzcd}
\lstick{4}   &  \gate{H} &\qw \\
\end{tikzcd}$ & \setlength\arraycolsep{0mm} \renewcommand\arraystretch{0.75} $\begin{matrix}
              &X &I &I &I &I \\
              &I &X &Z &I &Z \\
              &I &Z &X &I &I \\
              &I &Z &I &X &X \\
              &I &Z &I &I &X
               \end{matrix}$ &   $\ket{+0+++} + \ket{+1-+-}$   \\ \hline
$rowsum(5 \rightarrow 4)$  & \setlength\arraycolsep{0mm} \renewcommand\arraystretch{0.75} $\begin{matrix}
              &X &I &I &I &I \\
              &I &X &Z &I &Z \\
              &I &Z &X &I &I \\
              &I &I &I &X &I \\
              &I &Z &I &I &X
               \end{matrix}$ &   $\ket{+0+++} + \ket{+1-+-}$   \\ \hhline{|= = =|}
              \multicolumn{1}{|c|}{Final Graph} &  \multicolumn{2}{c|}{ \parbox[c]{2.8cm}{\includegraphics[height=2.3cm]{Ctrl-v_graph_00.pdf}}} \\ \hline
\end{tabular}}

\caption{Graph conversion of the final state in Tab.~\ref{tb:ctrl-v_clifford}. Input qubits are colored green and output qubits are colored blue.}
\label{tb:ctrl-v_graph}
\end{minipage}
\end{table*}
\begin{table*}
\begin{minipage}[t]{.45\textwidth}
\resizebox{0.9\textwidth}{!}{%
\begin{tabular}[t]{|c|c|c|} \hline
\setlength\arraycolsep{0mm}
    Operation & Stabilizer & State  \\ \hline
                    & \setlength\arraycolsep{0mm} \renewcommand\arraystretch{0.75} $\begin{matrix}
              &X &I &I &I &I \\
              &I &Z &I &I &I \\
              &I &I &Z &I &I \\
              &I &I &I &Z &I \\
              &I &I &I &I &Z
               \end{matrix}$ &  $\ket{+0000}$ \\ \hline
    $ \begin{tikzcd}
\lstick{1}   &  \ctrl{1} &\qw \\
\lstick{3}  & \targ{}  &\qw \\
\end{tikzcd}$ & \setlength\arraycolsep{0mm} \renewcommand\arraystretch{0.75} $\begin{matrix}
              &X &I &X &I &I \\
              &I &Z &I &I &I \\
              &Z &I &Z &I &I \\
              &I &I &I &Z &I \\
              &I &I &I &I &Z
               \end{matrix}$ &  $\ket{00000} + \ket{10100}$ \\ \hline
    $ \begin{tikzcd}
\lstick{2}   &  \gate{H} &\qw \\
\end{tikzcd}$ & \setlength\arraycolsep{0mm} \renewcommand\arraystretch{0.75} $\begin{matrix}
              &X &I &X &I &I \\
              &I &X &I &I &I \\
              &Z &I &Z &I &I \\
              &I &I &I &Z &I \\
              &I &I &I &I &Z
               \end{matrix}$ &  \makecell{\renewcommand\cellalign{cc}  $\ket{00000} + \ket{01000}$ \\ $\ket{10100} + \ket{11100}$  } \\ \hline
    $ \begin{tikzcd}
\lstick{2}   &  \ctrl{1} &\qw \\
\lstick{3}  & \targ{}  &\qw \\
\end{tikzcd}$ & \setlength\arraycolsep{0mm} \renewcommand\arraystretch{0.75} $\begin{matrix}
              &X &I &X &I &I \\
              &I &X &X &I &I \\
              &Z &Z &Z &I &I \\
              &I &I &I &Z &I \\
              &I &I &I &I &Z
               \end{matrix}$ &  \makecell{\renewcommand\cellalign{cc}  $\ket{00000} + \ket{01100}$ \\ $\ket{10100} + \ket{11000}$  } \\ \hline
$\begin{tikzcd}
\lstick{3}   &  \ctrl{1} &\qw \\
\lstick{4}  & \targ{}  &\qw \\
\lstick{2}  &  \ctrl{1} &\qw \\
\lstick{5}  & \targ{}  &\qw \\
\end{tikzcd}$ & \setlength\arraycolsep{0mm} \renewcommand\arraystretch{0.75} $\begin{matrix}
              &X &I &X &X &I \\
              &I &X &X &X &X \\
              &Z &Z &Z &I &I \\
              &I &I &Z &Z &I \\
              &I &Z &I &I &Z
               \end{matrix}$ &  \makecell{\renewcommand\cellalign{cc} $\ket{00000} + \ket{01111}$ \\ $\ket{10110} + \ket{11001}$  } \\ \hline
$ \begin{tikzcd}
\lstick{4}   &  \targ{} &\qw \\
\lstick{5}  & \ctrl{-1}  &\qw \\
\end{tikzcd}$ &  \setlength\arraycolsep{0mm}\renewcommand\arraystretch{0.75} $\begin{matrix}
              &X &I &X &X &I \\
              &I &X &X &I &X \\
              &Z &Z &Z &I &I \\
              &I &I &Z &Z &Z \\
              &I &Z &I &I &Z
               \end{matrix}$ &  \makecell{\renewcommand\cellalign{cc} $\ket{00000} + \ket{01101}$ \\ $\ket{10110} + \ket{11011}$ } \\ \hline
$\begin{tikzcd}
\lstick{5}   &  \gate{H} &\qw \\
\end{tikzcd}$ & \setlength\arraycolsep{0mm} \renewcommand\arraystretch{0.75} $\begin{matrix}
              &X &I &X &X &I \\
              &I &X &X &I &Z \\
              &Z &Z &Z &I &I \\
              &I &I &Z &Z &X \\
              &I &Z &I &I &X
               \end{matrix}$ &  \makecell{\renewcommand\cellalign{cc} $\ket{0000+} + \ket{0110-}$ \\ $\ket{1011+} + \ket{1101-}$ }  \\ \hline
\end{tabular}}
\caption{Gate transformations of the controlled-$V^{\dagger}$ decomposition give in figure~\ref{Fig:inverse_icm_example} excluding final measurements for inputs $\ket{ab} = \ket{+0}$.}
\label{tb:ctrl-v_clifford_+0}
\end{minipage}%
\begin{minipage}[t]{.45\textwidth}
\resizebox{0.9\textwidth}{!}{%
\begin{tabular}[t]{|c|c|c|} \hline
    Operation & Stabilizer & State  \\ \hline
                    & \setlength\arraycolsep{0mm} \renewcommand\arraystretch{0.75} $\begin{matrix}
              &X &I &X &X &I \\
              &I &X &X &I &Z \\
              &Z &Z &Z &I &I \\
              &I &I &Z &Z &X \\
              &I &Z &I &I &X
               \end{matrix}$ &  \makecell{\renewcommand\cellalign{cc} $\ket{0000+} + \ket{0110-}$ \\ $\ket{1011+} + \ket{1101-}$}   \\ \hline
    $ \begin{tikzcd}
\lstick{3}   &  \gate{H} &\qw \\
\end{tikzcd}$ & \setlength\arraycolsep{0mm} \renewcommand\arraystretch{0.75} $\begin{matrix}
              &X &I &Z &X &I \\
              &I &X &Z &I &Z \\
              &Z &Z &X &I &I \\
              &I &I &X &Z &X \\
              &I &Z &I &I &X
               \end{matrix}$ &  \makecell{\renewcommand\cellalign{cc} $\ket{00+0+} + \ket{01-0-}$ \\ $\ket{10-1+} + \ket{11+1-}$} \\ \hline
$rowsum(3 \rightarrow 4)$ & \setlength\arraycolsep{0mm} \renewcommand\arraystretch{0.75} $\begin{matrix}
              &X &I &Z &X &I \\
              &I &X &Z &I &Z \\
              &Z &Z &X &I &I \\
              &Z &Z &I &Z &X \\
              &I &Z &I &I &X
               \end{matrix}$ &  \makecell{\renewcommand\cellalign{cc} $\ket{00+0+} + \ket{01-0-}$ \\ $\ket{10-1+} + \ket{11+1-}$} \\ \hline
    $ \begin{tikzcd}
\lstick{4}   &  \gate{H} &\qw \\
\end{tikzcd}$ & \setlength\arraycolsep{0mm} \renewcommand\arraystretch{0.75} $\begin{matrix}
              &X &I &Z &Z &I \\
              &I &X &Z &I &Z \\
              &Z &Z &X &I &I \\
              &Z &Z &I &X &X \\
              &I &Z &I &I &X
               \end{matrix}$ &  \makecell{\renewcommand\cellalign{cc} $\ket{00+++} + \ket{01-+-}$ \\ $\ket{10--+} + \ket{11+--}$} \\ \hline
$rowsum(5 \rightarrow 4)$ & \setlength\arraycolsep{0mm} \renewcommand\arraystretch{0.75} $\begin{matrix}
              &X &I &Z &Z &I \\
              &I &X &Z &I &Z \\
              &Z &Z &X &I &I \\
              &Z &I &I &X &I \\
              &I &Z &I &I &X
               \end{matrix}$ &  \makecell{\renewcommand\cellalign{cc} $\ket{00+++} + \ket{01-+-}$ \\ $\ket{10--+} + \ket{11+--}$}  \\ \hhline{|= = =|}
              \multicolumn{1}{|c|}{Final Graph} &  \multicolumn{2}{c|}{ \parbox[c]{3.5cm}{\includegraphics[height=3cm]{Ctrl-v_graph_+0.pdf}}} \\ \hline
\end{tabular}}
\caption{Graph conversion of the final state in Tab.~\ref{tb:ctrl-v_clifford_+0}}
\label{tb:ctrl-v_graph_+0}
\end{minipage}
\end{table*}

\onecolumngrid 

\clearpage

\section{Graph conversion algorithm} \label{apx:graph_conversion}

\begin{algorithm}
\DontPrintSemicolon
 \SetKwInOut{Input}{input}
 \SetKwInOut{Output}{output}
 \SetKwFunction{Func}{ToGraph}
 \SetKwFunction{ApplyH}{H}
 \SetKwFunction{ApplyZ}{Z}
 \SetKwFunction{ApplyPdag}{P$^\dag$}
 \SetKwFunction{rowswap}{rowswap}
 \SetKwFunction{rowadd}{rowadd}
 \Input{Tableau $X$, $Z$, $P$.}
 \Output{Adjacency matrix $A$.}
 \BlankLine
 \Begin{
  \tcc{Make $X$-block full rank, $O(n^3)$}
  \For{$i\leftarrow 1$ \KwTo $n$}{
    \tcc{How many $X$'s are in the lower-triangular part of the $i$th column?}
    $\sigma=\sum_{k=i}^n X_{k,i}$\;
    \tcc{If there aren't any, pull them over from the $Z$-block}
    \If{$\sigma=0$}{
        \ApplyH{$i$}
    }
    \tcc{Perform any necessary row-swap to bring a leading $X$ onto the diagonal}
    \For{$j\leftarrow i$ \KwTo $n$}{
        \If{$X_{j,i}=1$}{
            \rowswap{$i\leftrightarrow j$}\;
            \textbf{break}\;
        }
    }
    \tcc{Eliminate $X$'s below the diagonal}
    \For{$j\leftarrow i+1$ \KwTo $n$}{
        \If{$X_{j,i}=1$}{
            \rowadd{$i\rightarrow j$}\;
        }
    }
  }
  \tcc{Diagonalize $X$-block, $O(n^3)$}
  \For{$i\leftarrow n-1$ \KwTo $1$}{
    \For{$j\leftarrow n$ \KwTo $i+1$}{
        \If{$X_{i,j}=0$}{
            \rowadd{$j\rightarrow i$}
        }
    }
  }
  \tcc{Make $Z$-block diagonals zero and correct phases, $O(n)$}
  \For{$i\leftarrow 1$ \KwTo $n$}{
    \If{$Z_{i,i}=1$}{
    	\ApplyPdag{$i$}
    }
    \If{$P_i=-1$}{
        \ApplyZ{$i$}
    }
    
  }
  \tcc{$Z$-block contains adjacency matrix}
  \Return{$A\leftarrow Z$}
 }
\caption{The graph conversion algorithm, with $O(n^3)$ runtime.}
\end{algorithm}

\end{document}